\documentclass[11pt]{article}

\usepackage{pspicture}

\title{\huge Conformal Sequestering Simplified}

\author{\\ {\Large Martin Schmaltz} \\ \\
        Physics Department, Boston University \\ 
        590 Commonwealth Ave, Boston, MA 02215 \\
        E-mail: {\rm \tt schmaltz@bu.edu}\\ \\ \\
        {\Large Raman Sundrum}\\ \\
        Department of Physics and Astronomy, Johns Hopkins University \\
        3400 North Charles St., Baltimore, MD 21218 \\
        E-mail: {\rm \tt sundrum@pha.jhu.edu}\\ \\}

\newcommand{\eqn}[1]{\label{eq:#1}}
\newcommand{\refeq}[1]{(\ref{eq:#1})}

\newcommand{\beq}{\begin{eqnarray}}
\newcommand{\eeq}{\end{eqnarray}}

\newcommand\gsim{\mathrel{%
   \rlap{\raise 0.511ex \hbox{$>$}}{\lower 0.511ex \hbox{$\sim$}}}}
\newcommand\lsim{\mathrel{
   \rlap{\raise 0.511ex \hbox{$<$}}{\lower 0.511ex \hbox{$\sim$}}}}

\newcommand{\drawsquare}[2]{\hbox{%
\rule{#2pt}{#1pt}\hskip-#2pt
\rule{#1pt}{#2pt}\hskip-#1pt
\rule[#1pt]{#1pt}{#2pt}}\rule[#1pt]{#2pt}{#2pt}\hskip-#2pt
\rule{#2pt}{#1pt}}

\newcommand{\Yfund}{\raisebox{-.5pt}{\drawsquare{6.5}{0.4}}}
\newcommand{\ignore}[1]{}

\begin{document}

\baselineskip=17pt 
\pagestyle{plain} 

\begin{titlepage} 

\maketitle

\vskip.5in

\abstract{Sequestering is important for obtaining flavor-universal soft
masses in models where supersymmetry breaking is mediated at high scales.
We construct a simple and robust class of hidden sector
models which sequester themselves from the visible sector
due to strong and conformally invariant hidden dynamics.
Masses for hidden matter eventually break the conformal symmetry and
lead to supersymmetry breaking by the mechanism recently discovered by
Intriligator, Seiberg and Shih.
We give a unified treatment of subtleties due to global symmetries
of the CFT. There is enough review for the paper to constitute
a self-contained account of conformal sequestering.}


\thispagestyle{empty} 
\setcounter{page}{0} 
\end{titlepage}

\section{Introduction}

Several of the effective field theory 
solutions to the flavor problem
of weak scale SUSY 
rely on ``sequestering'' of the hidden sector from
the visible sector,  in order to avoid flavor-violating 
squark and slepton masses from operators of the form
\beq
\label{thebadguy}
\int d^4 \theta\ c_{ij} \frac{\Phi_i^\dagger \Phi_j X^\dagger X}{M_{Pl}^2}.
\eeq
Here, $\Phi_i$ stands for the $i$'th generation of an MSSM
quark or lepton superfield, $X$ is a hidden sector superfield 
with a SUSY breaking $F_X$-component and $M_{Pl}$ is the 
Planck scale. This operator arises from integrating out
heavy (string) physics near the Planck scale and is expected
to have coefficients, $c_{ij} \sim {\cal O}(1)$, which are flavor-violating
 because  Yukawa couplings require breaking of flavor symmetries.

In order for the flavor-preserving scalar masses generated from
anomaly-mediated SUSY breaking (AMSB) \cite{RS0,AMSB2},
gaugino-mediation \cite{gaugino1, gaugino2}, or any other such high-scale
mediation mechanism \cite{gravloops1,gravloops2,mirage1,mirage2,mirage3,mirage4}
to dominate, the operator of Eq.~(\ref{thebadguy})
must be suppressed. For example, the (flavor-universal) 
contributions to visible scalar masses from AMSB are loop-factor suppressed 
relative to those from the above direct coupling to the hidden sector, 
\begin{eqnarray}
m_{AMSB}^2 &\sim& 
 \left(\!\frac{g_{SM}^2}{16 \pi^2}\!\right)^{\!2} \frac{|F_X|^2 }{M_{Pl}^2} 
\end{eqnarray}
\begin{eqnarray}
\eqn{direct}
(m_{direct}^2)_{ij} &=& c_{ij}\, \frac{|F_X|^2}{M_{Pl}^2}\ \ .
\end{eqnarray}
This implies that Eq.~(\ref{thebadguy}) must be suppressed by at least
$\mathcal O(10^{-6} - 10^{-7})$ for AMSB to
solve the flavor problem. Sequestering refers to 
this suppression, even beyond Planck-suppression,
 of direct hidden-visible couplings.%

Conformal sequestering \cite{LS1,LS2} accomplishes this suppression 
 by strong-coupling hidden sector anomalous dimensions
(or alternatively by large visible-sector anomalous dimensions \cite{NS1,NS2})
in the running of dangerous operators such as Eq.~(\ref{thebadguy}) 
from the Planck scale down to the SUSY-breaking intermediate scale, 
\beq
\Lambda_{int} \equiv \sqrt{F_X} \sim \ignore{3 \times} 10^{11}\ {\rm GeV}\ .
\eeq 
A  virtue of conformal sequestering is that it depends on
purely four-dimensional, renormalizable dynamics, 
rather than  non-renormalizable extra-dimensional effective field theories
as originally proposed \cite{RS0}. The robustness 
and plausibility of sequestering can therefore be addressed 
with less questionable assumptions about string theory ultraviolet completions
\cite{Dine1,Kachru}. 
This is still a non-trivial task because of the key role played by
non-perturbative strong dynamics, and progress depends on 
inferences based on global symmetries, SUSY and field theory dualities. 
The central obstacle for conformally sequestered models of SUSY breaking 
is that the symmetries used to understand strongly-coupled SUSY 
dynamics also yield conserved currents with vanishing, rather than
the requisite large, anomalous dimensions \cite{NS1,NS2,LS1,LS2}.
Nevertheless successful models have been constructed.
While early models were somewhat complicated \cite{LS1,LS2}, technology
has improved and more plausible models with conformal sequestering 
now exist \cite{Yanagida1,Yanagida2}.

In this paper, we present a particularly simple class of hidden sector models 
which achieve conformal sequestering suitable for AMSB by taking advantage
of the SUSY breaking metastable vacua of
supersymmetric QCD recently discovered by Intriligator, Seiberg and Shih (ISS)
\cite{ISS}. The simplicity of our models paints a
highly plausible picture of how anomaly-mediation%
\footnote{In this paper we are using AMSB as an example for a flavor-blind
high scale mediation mechanism. Of course, fully realistic mediation mechanisms
(with positive slepton masses) require additional structure \cite{gaugomaly}, but our
focus here is on the issues of conformal sequestering and hidden sector
SUSY breaking.}
might dominate weak scale SUSY breaking. 
We also provide a broader perspective and collect general results on model
building. Most of these have appeared previously in the
literature \cite{NS1,NS2,LS1,LS2,K1,K2,K3,Dine2,Yanagida1,Yanagida2},
but the arguments of Sections 4.2 - 4.3 on ``exact flavor currents'' and
Section 4.6. on emergent symmetries are new.
We have tried to incorporate enough review material to make this paper 
a self-contained introduction to conformal sequestering.

Here is the basic plot of conformal sequestering. Hidden 
sector dynamics is assumed to be in the vicinity of a strongly-coupled 
superconformal fixed point over a large enough hierarchy 
between the $M_{Pl}$ and 
$\Lambda_{int}$ so that strong running effects can 
suppress dangerous hidden-visible couplings. 
We focus on the resulting
 visible {\it scalar} mass-squareds because other soft terms 
can be protected by chiral symmetries. We work in flat spacetime even 
though AMSB relies on supergravity since we are only addressing the issue of 
sequestering, the suppression of the unwanted ``background'' to AMSB.
Since at strong coupling it is not obvious which operators are 
the most relevant and therefore most dangerous, we discuss sequestering
of a general hidden sector operator ${\cal O}_{hid}$.
Denoting the scalar component of $\Phi$ by $\phi$ and, for simplicity, working with
the component Lagrangian, we have
\begin{eqnarray}
\eqn{eff}
\Delta {\cal L}_{hid-vis}(M_{Pl}) &\sim& \frac{1}{M_{Pl}^{n - 2}} \
\phi^{\dagger} \phi \ {\cal O}_{hid} \nonumber \\
\Delta {\cal L}_{hid-vis}(\mu) &\sim& \left(\!\frac{\mu}{M_{Pl}}\!\right)^{\gamma}\,
\frac{1}{M_{Pl}^{n - 2}}\ \phi^{\dagger} \phi\  {\cal O}_{hid}\ ,
\end{eqnarray}
where $n$ and $\gamma$ are the canonical and anomalous dimensions of ${\cal O}_{hid}$
respectively. Here we neglect the 
presumably weak visible running effects.
If this strong running holds down to energies not far above $\Lambda_{int}$, 
then  (except for the AMSB contribution), 
\begin{equation}
m_{\phi}^2 \sim \frac{\Lambda_{int}^4}{M_{Pl}^2} \,
\left(\!\frac{\Lambda_{int}}{M_{Pl}}\!\right)^{n + \gamma - 4},
\end{equation}
which is a suppression of the direct contribution, Eq.~\refeq{direct}, by
an additional  $(\frac{\Lambda_{int}}{M_{Pl}})^{n + \gamma - 4}$. This is 
sufficient for solving the flavor problem provided 
\begin{equation}
\label{one}
n + \gamma - 4 \gsim 1
\end{equation}
for all hidden sector operators ${\cal O}_{hid}$. 

What kind of fixed point 
satisfies this condition for all Lorentz-invariant operators? 
When asking about visible masses, it is sufficient to 
think of $\phi^{\dagger} \phi$ at zero momentum, that is as a spacetime 
constant. In this case, as far as the hidden dynamics in the conformal 
regime is concerned, Eq.~\refeq{eff} is just adding the interaction ${\cal O}_{hid}$ to the 
fixed point with a small $\phi$-dependent coupling.
If the fixed point is IR-attractive, then all 
such perturbations should be irrelevant, so that the scaling dimension of 
${\cal O}_{hid}$, $n + \gamma$, is larger than 4. 
At a strongly-coupled fixed point with no small numbers, irrelevant 
couplings are ${\cal O}(1)$ irrelevant, which is the 
condition, Eq. (\ref{one}). Thus, we must simply arrange for 
the hidden sector dynamics to be close to a strongly-coupled 
IR-attractive fixed-point below the Planck scale in order to sequester.
However, as mentioned above, there is an important subtlety involving global symmetries
which we treat carefully in this paper.

Ultimately the hidden sector must break SUSY at $\Lambda_{int}$, 
and thus must violate conformal symmetry above this scale.
We introduce hidden masses $m_X \gsim \Lambda_{int}$, below which the
hidden dynamics triggers 
spontaneous 
SUSY breaking. The relevant scales are summarized in Figure 1.

\begin{figure}[htb]
\hskip1.4in
\begin{picture}(150,140)

\linethickness{2pt}

\put(50,-10){\vector(0,1){150}}

\put(45,120){\line(1,0){10}}
\put(45,45){\line(1,0){10}}
\put(45,20){\line(1,0){10}}

\linethickness{1pt}

\put(5,20){$\Lambda_{int}$}
\put(5,45){$m_X$}
\put(5,120){$M_{Pl}$}

\put(70,90){conformal dynamics}
\put(70,45){conformal symmetry breaking}
\put(70,20){SUSY breaking}

\end{picture}
\caption{Scales and dynamics.}
\end{figure}
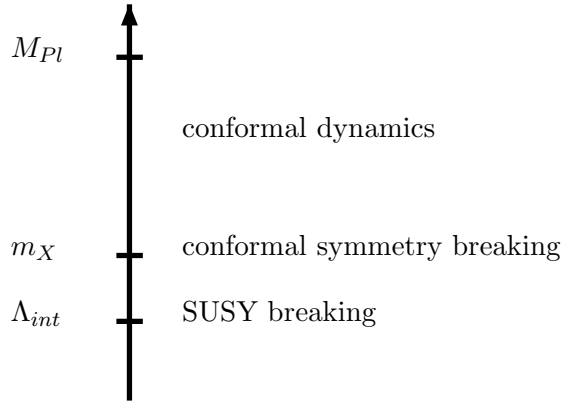

This paper is organized as follows. In Section 2, we give a quick 
 introduction to conformal sequestering within two toy models, 
without reference to SUSY breaking.  
The first toy model is non-supersymmetric, giving the
simplest concrete illustration of conformal sequestering. 
The second toy model is
just SUSY QCD, and we give a new streamlined discussion of
 conformal sequestering there, with some back-up material reviewed 
in an Appendix. 
In Section 3, we introduce our new 
class of strongly coupled hidden sector models, 
explain how the UV
conformal regime gives way to SUSY breaking in the IR, and briefly 
indicate how conformal sequestering is accomplished as far as terms of the 
form of Eq.~(\ref{thebadguy}) are concerned.   In Section 4, 
we carefully examine the various aspects of 
sequestering in our new
 class of models, including subtleties previously neglected.  
In Section 5, we 
broaden the considerations of Sections 3 and 4 
to a set of model-building rules, 
both for the purpose of 
constructing new models as well as for assessing the general 
plausibility of sequestering and the high-scale mediation  mechanisms that 
depend on it. 

The casual reader, wanting a quick acquaintance with our new models and a 
summary of how they perform, need only read the Introduction and Section 3.

\section{Toy Models of Conformal Sequestering}

In this Section we use two simple examples to demonstrate
conformal sequestering. The first is a scalar theory with a quartic
coupling in $4\!-\!\epsilon$ dimensions. It has a non-trivial infrared (IR) attractive
fixed point at weak coupling which allows us to give explicit formulae in perturbation
theory.
The theory is also non-supersymmetric, which makes it clear that conformal sequestering
is a property of conformal theories, not just SUSY. Our second example is
supersymmetric QCD in four dimensions in the conformal window.

\subsection{Non-supersymmetric example}

Consider a single massless real scalar field $\hat X$ with a
quartic interaction in $4-\epsilon$ dimensions
\beq
\mathcal L = \frac12 \left( \partial_\mu \hat X\right)^2 - \mu^\epsilon \frac{\lambda}{4!} \hat X^4 \ .
\eeq
We gave the field $\hat X$ a hat as a reminder that $\hat X$ has a canonically normalized
kinetic term and we factored out the explicit factor of $ \mu^\epsilon$ from the coupling
constant to make $\lambda$ dimensionless. With $\mu^\epsilon$ factored out,
the Lagrangian is invariant under
conformal transformations iff $\lambda$ is scale invariant ($\mu$ independent).

The $\overline{\rm MS}$ beta function for $\lambda$ at one loop is
\beq
\beta(\lambda) = \mu \frac{\partial \lambda}{\partial \mu} - \epsilon \lambda + 3 \frac{\lambda^2}{16 \pi^2} 
\eeq
The first term is the ``classical scaling'', it comes from the explicit factor of
$\mu^\epsilon$ in the interaction in $4\!-\!\epsilon$ dimensions.
The second term is the one-loop correction.
The beta function vanishes for $\lambda=\lambda_*= \frac{16 \pi^2 \epsilon}{3}$
and the theory is conformal. To study the theory near conformality we expand
the beta function near the fixed point to linear order
\beq
\beta(\lambda) \simeq 0 +
\left. \frac{\partial \beta}{\partial \lambda} \right|_{\lambda=\lambda_*}\!\!\!\! (\lambda-\lambda_*)
=\epsilon\, (\lambda-\lambda_*)
\eqn{linearbeta}
\eeq
and integrate to obtain
\beq
\lambda(\mu)=\lambda_* + \left(\frac{\mu}{M}\right)^\epsilon (\lambda(M)-\lambda_*)
\eqn{toyrunningcoupling}
\eeq
We see that the fixed point is attractive and that deviations from the fixed
point are scaled away by a factor of $(\mu/M)^\epsilon$, that is
 irrelevant by $\epsilon$, despite the fact that the coupling is canonically 
``relevant'' by $\epsilon$. 

We used perturbation theory to compute the beta function and establish
the existence of a fixed point. At strong coupling we cannot prove the existence
of the fixed point, but if we assume that an isolated IR fixed point exists,
we can expand the beta function near the fixed point and derive
power law running,  corresponding to  deviations 
from the fixed point being irrelevant  by 
$\left. \partial \beta/\partial \lambda\right|_{\lambda=\lambda_*} \equiv {\beta'}_{\!*} $.

To set the stage for what comes later, let us translate these results into a different
basis reached by rescaling the field $\hat X$ to make the coupling constant equal to
$\lambda_*$ at all energy scales ($\lambda_*$ is defined as the fixed point coupling
in the canonical basis). The kinetic term is now multiplied by a $Z$ factor
\beq
\mathcal L = \frac12 Z(\mu) ( \partial_\mu X )^2
- \mu^\epsilon \frac{\lambda_*}{4!} X^4 \ ,
\eeq
which accounts for all non-trivial running.
We determine $Z(\mu)$ from \refeq{toyrunningcoupling} by rescaling
$X = (\lambda(\mu)/\lambda_* )^{1/4}\, \hat X$
to obtain
\beq
Z(\mu)=\sqrt{\frac{\lambda_*}{\lambda}}
&=& \left[\frac{1}{1+(\frac{\mu}{M})^\epsilon(\frac{\lambda(M)}{\lambda_*}-1)}\right]^{1/2} \\
&=& \left[\frac{1}{1+(\frac{\mu}{M})^\epsilon(Z^{-2}_M-1)}\right]^{1/2}
\simeq 1 + \left(\frac{\mu}{M}\right)^\epsilon \left(Z_M-1\right)
\eqn{toyscaling}
\eeq
where in the second line we traded the UV boundary value of the coupling $\lambda(M)$
for the corresponding UV value of the wave function $Z_M \equiv \sqrt{\lambda_*/\lambda(M)}$,
and approximated for $Z_M$ close to 1. 
In this basis, the approach to the fixed point is $Z(\mu)\rightarrow 1$
when $\mu \rightarrow 0$; deviations from the fixed point are again seen to be 
irrelevant by $\epsilon$.

Let us now discuss sequestering by imagining that our toy model is the ``hidden sector''
coupled to a ``visible sector'' represented by a free real scalar field $\phi$.
We assume that the higher dimensional operator
\beq
c\, \frac{\phi^2}{M^2}\  \frac12 (\partial_\mu X )^2
\eqn{toyoperator}
\eeq
couples the two sectors at the UV scale $M$. Note that there is no hat on the $X$ because
we are using the basis discussed in the previous paragraphs.

We now show that this operator sequesters, i.e.
it obtains an anomalous dimension from hidden sector dynamics which increases its
scaling dimension so that it becomes more irrelevant in the IR than the naive $1/M^2$.
One can compute the anomalous dimension directly by computing Feynman diagrams in which
the operator is dressed with $X$ loops but it is easier to obtain the result with a trick
\cite{A-HGLR}.

The argument uses the fact that the visible sector fields are background fields in this
calculation (the running of the operator due to visible sector interactions is negligible
compared to hidden sector running), and we may choose a simple configuration for $\phi$.
The most convenient choice is a constant $\phi=\phi_0 \ll M$. Then, from the point
of view of the hidden sector dynamics, the operator \refeq{toyoperator} becomes a small
additional contribution to the kinetic term for $X$. More precisely, $Z_M$ gets a
contribution equal to $c\, \phi_0^2/M^2$.
But we know how the $X$ kinetic term is renormalized from equation \refeq{toyscaling},
any dependence on $Z_M$ and therefore any dependence on $c\, \phi_0^2/M^2$ scales 
away with a factor of $(\mu/M)^{\epsilon}$. This scaling is valid for any constant
$\phi$ and therefore also for the operator
\beq
\left(\frac{\mu}{M}\right)^\epsilon c \frac{\phi^2}{M^2}\ \ \frac12 (\partial_\mu X )^2 \ .
\eeq

The essence of the argument is that for constant $\phi$ the operator \refeq{toyoperator}
can be thought of as a contribution to the UV boundary conditions for a coupling in the
hidden sector theory. But since the hidden sector runs towards an IR fixed point, the 
theory ``forgets'' the UV values of the coupling constants. In the IR, the coupling constants
depend only on the CFT, not on the UV boundary conditions. Numerically, the ``forgetting'' 
proceeds by power law scaling. Therefore the operator  \refeq{toyoperator} must scale away
with a power of $\mu/M$.
This argument does not depend on perturbation theory,
it relies only on the assumption that we have an IR attractive fixed point.
In the strongly coupled case the exponent
would be ${\beta'}_{\!*}$. It should also be clear that the argument
generalizes to CFTs with multiple couplings. Again, UV perturbations to an IR 
attractive fixed point, due to the visible sector, simply scale away. 


\subsection{Supersymmetric QCD}

While running $Z$-factors are unusual in a discussion of simple scalar
field theories they are very natural in supersymmetric theories.
In our second example \cite{LS1,LS2}, supersymmetric QCD in the conformal window,
all running except
for a one-loop contribution to gauge couplings occurs in $Z$-factors when working
in a holomorphic basis. As in the example discussed above, it is most convenient to
work in a basis where all running (including the one-loop running of the holomorphic gauge
coupling) has been scaled into $Z$-factors for the matter fields. This is
accomplished by performing a ``Konishi-anomalous'' \cite{Konishi}
transformation of the quark fields.
The Lagrangian for SUSY QCD is then
\beq
\mathcal L = R(\mu) \int d^4\theta \  (Q^\dagger Q + \bar Q^\dagger \bar Q)
+ \int d^2\theta \  W_\alpha W^\alpha  + h.c. 
\eeq
where $R(\mu)$ is the running $Z$-factor. Near the fixed point ($R\rightarrow R_*=const.$) one finds
(see \cite{LS1} and the Appendix)
\beq
R(\mu)/R_*=\left[R(M)\right]^{(\frac{\mu}{M})^{{\beta'}_{\!*}}} \simeq
1+\left(\frac{\mu}{M}\right)^{{\beta'}_{\!*}} (R(M)-1)
\eqn{sqcdseq}
\eeq
As in the previous example, deviations from the fixed point scale away with a
power of $\mu/M$. ${\beta'}_{\!*}$ is the derivative of the
$\beta$ function for the gauge coupling at the fixed point,
it is not calculable at strong coupling but is expected to be of order 1.

The operator we wish to sequester with conformal dynamics is of the form
\beq
\int d^4\theta \ c \frac{\Phi^\dagger \Phi}{M^2} (Q^\dagger Q +\bar Q^\dagger \bar Q) \ .
\eqn{susyop}
\eeq
Again, by using the trick of considering constant (scalar) background values for the field
$\Phi$  so that $c \frac{\Phi^\dagger \Phi}{M^2}$ becomes a number which can be absorbed
into $R(M)$, we can derive the sequestering from the scaling \refeq{sqcdseq},
\beq
\int d^4\theta \ \left(\frac{\mu}{M}\right)^{\gamma}
c \frac{\Phi^\dagger \Phi}{M^2} (Q^\dagger Q +\bar Q^\dagger \bar Q), 
\eeq
where the anomalous dimension associated to small 
$Z$-shifts (the kinetic operator) is $\gamma = {\beta'}_{\!*}$.

\section{A Self-sequestering Hidden Sector}

In this Section we present a hidden sector model which exhibits
conformal sequestering and spontaneous SUSY breaking.
We use this model as a concrete example because of its simplicity. 
Similar models are straightforward to construct.

The model is $\mathcal{N}$=1 supersymmetric, with gauge group $SU(N)$ and
$F$ chiral superfield ``flavors" $Q, \overline Q$ in the fundamental and anti-fundamental
representations and an adjoint $A$. 


\begin{equation}
\begin{array}{c|c|cc}
& SU(N) & SU(F) & SU(F) \\
\hline
Q & \Yfund  & \Yfund & 1 \\
\bar{Q} & \overline{\Yfund} & 1 & \overline{\Yfund} \\
A & adj & 1 & 1 \\
\end{array}
\end{equation}

For the range of flavors of interest, $N<F<\frac32 N$, the 
$SU(N)$ gauge coupling is asymptotically free and grows in the infrared.
In order to reduce the number of allowed couplings we
weakly gauge the diagonal ``vector'' $SU(F)$  group. We choose this
gauge coupling to be weak so that we can ignore its effects on the
dynamics.
Our superpotential is
\beq
W=m_Q\, Q\overline Q + \frac{m_A}{2} A^2 + \frac{\kappa}{3} A^3
\eeq
where the scales are arranged as shown in Figure 2,
$m_Q<m_A\ll \Lambda_{\kappa}$$ \,\lsim $$\,\Lambda_N$.
This model was studied extensively in the context of supersymmetric duality
in \cite{K,KS,KSS}.

\begin{figure}[htb]
\hskip1.4in
\begin{picture}(280,225)

\linethickness{2pt}

\put(53,13){\vector(0,1){198}}

\put(48,170){\line(1,0){10}}
\put(48,150){\line(1,0){10}}
\put(48,55){\line(1,0){10}}
\put(48,30){\line(1,0){10}}

\linethickness{1pt}

\put(26,40){\vector(1,0){20}}
\put(5,40){$\Lambda_{int}$}
\put(28,132){\vector(1,0){20}}
\put(5,130){$M_{Pl}$}

\put(65,170){$\Lambda_{N}$}
\put(65,150){$\Lambda_{\kappa}$}
\put(65,55){$m_A$}
\put(65,30){$m_Q$}

\put(46,220){$UV$}
\put(46,0){$IR$}

\put(100,200){asymptotic freedom}
\put(100,160){first CFT}
\put(100,110){second CFT,}
\put(100,95){sequestering}
\put(100,40){SUSY breaking}

\end{picture}

\caption{Scales and dynamics of our model. 
$\Lambda_N$ and $\Lambda_\kappa$, respectively, are the scales at which the
hidden sector gauge coupling and Yukawa coupling $\kappa$ become strong
and the theory transitions to a conformal fixed point.
$m_A$ and $m_Q$ are masses which explicitly break conformal symmetry
and trigger SUSY breaking.
$M_{Pl}$ and $\Lambda_{int}$ are the Planck and intermediate scales.}
\end{figure}
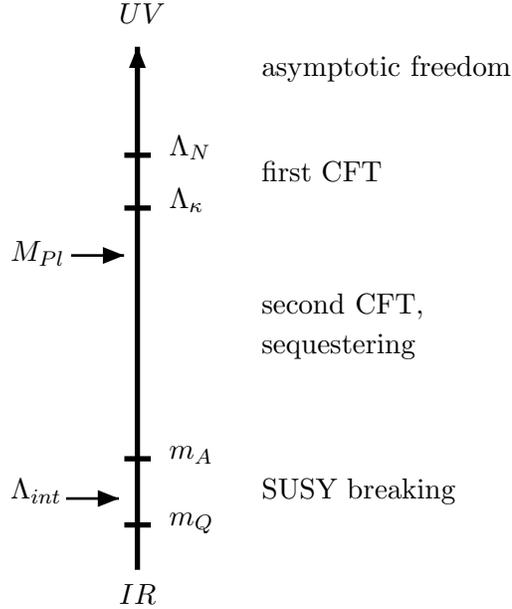

We will tell the story of the RG flow beginning above $\Lambda_N$, 
neglecting gravity, as a simplifying abstraction. Of course, 
in reality, (effective) field theory starts below $M_{Pl}$, and 
$\Lambda_N, \Lambda_{\kappa}$ are only apparent scales, mere theoretical 
crutches.
Above $\Lambda_N$, the theory is weakly coupled
and asymptotically free in the UV. At the scale $\Lambda_N$ the $SU(N)$
gauge coupling becomes strong and drives the theory towards a strongly
coupled fixed point. At this fixed point the operator $\kappa A^3$ is
relevant. Therefore $\kappa$ grows quickly and near $\Lambda_{\kappa}$
the theory approaches a new infrared fixed point at which both
the $SU(N)$ gauge coupling and $\kappa$ are strong.
Scaling dimensions near this fixed point are quite different from
the UV dimensions. For chiral primary operators they can
be determined from superconformal R-charges. For example,
$dim(Q\overline Q) = 3-2N/F$.
The dynamics remains governed by this fixed point until conformal symmetry is
broken by the mass of the adjoint field $A$ (determined by $m_A$)
and $A$ decouples from the infrared theory.
\footnote{The introduction of the hierarchy $m_X \ll \Lambda$ ``by 
hand'' may seem inelegant, but the small mass scales $m_X$  can naturally
arise from exponentially  small non-perturbative effects. A concrete example is 
gaugino condensation due to a pure super-Yang-Mills sector 
with Planck suppressed couplings in the superpotential
$
\int\! d^2 \theta\  {\cal W}_{\alpha}^2 \, (1 + \frac{X^2}{M_{Pl}^2})  \rightarrow 
\int\! d^2 \theta\ \Lambda_{SYM}^3\, (1 + \frac{X^2}{M_{Pl}^2} )
$
\ignore{\begin{equation}
{\cal L}_{SYM} = \int d^2 \theta (1 + {\cal O}(\frac{X^2}{M_{Pl}^2})) 
{\cal W}_{\alpha}^2  \rightarrow 
{\cal L}_{eff} = \int d^2 \theta \Lambda_{SYM}^3 (1 + 
{\cal O}(\frac{X^2}{M_{Pl}^2}) 
+ {\cal O}(\frac{X^4}{M_{Pl}^4})),  
\end{equation}
}
yielding a small mass term for $X$ with negligibly small 
 ${\cal O}(X^4/M_{Pl}^4)$ corrections.}

Just below $m_A$ the theory is SUSY QCD with a very strong gauge coupling
and too few flavors to remain conformal. Instead, $SU(N)$ charges are
confined and the theory flows to a free fixed point which is best
described in terms of dual degrees of freedom. This dual
has an $SU(F-N)$ gauge group and the particle content
\begin{equation}
\begin{array}{c|c|cc}
& SU(F-N) & SU(F) & SU(F) \\    \hline
q & \Yfund  & \overline{\Yfund} & 1 \\
\bar{q} & \overline{\Yfund} & 1&  \Yfund \\
M & 1  & \Yfund & \overline{\Yfund} \\ 
\end{array}
\end{equation}
Here $M$ is the operator map of the composite $SU(N)$ gauge invariant $Q\overline{Q}$
and $q$ and $\overline{q}$ are dual quarks. The dual has the superpotential
\beq
W=m^2 Tr M + M q \bar{q}
\eqn{dualsuperpot}
\eeq
where the mass scale $m^2 \propto m_Q$ is determined by matching the
scaling dimension of the operator $Q\overline Q$ in the UV to the
dimension of $Q\overline Q$ at the CFT fixed points and then to the
free field operator $M$ in the IR.
Using the flavor symmetries, holomorphy, and by considering various limits, one
can show that this superpotential is exact.

Finally, at energies of order $m$, the theory breaks SUSY with a
metastable ground state \cite{ISS}. To see this one
uses the fact that the theory is IR free so that
the kinetic terms for $M$, $q$, $\overline{q}$ are approximately canonical
and perturbation theory can be used to determine the vacuum structure.
Assuming small vacuum expectation values compared to the scale $m_A$,
the contribution to the potential from the F-term of $M$ is
\beq
V \sim \sum_{i,j} \left| m^2 \delta_{i j} + q_{i\alpha} \overline{q}_j^\alpha \right|^2 + \cdots
\eeq
where $i, j$ are $SU(F)$ flavor indices and $\alpha$ is an $SU(F-N)$ color index.
This potential is necessarily nonzero because $\delta_{i j}$ is a matrix of
rank $F$ whereas $q_{i\alpha}  {\overline {q}_j^\alpha}$ is at most of 
rank $F-N$ ($q$ is an $F\!-\!N \times F$ matrix). Thus SUSY is
spontaneously broken, and as in any O'Raifeartaigh model,
there is a classical flat direction. It corresponds to the scalar components of 
some of the diagonal elements of $M$ and is lifted once quantum corrections are taken into account.
The dominant effect comes from
one loop wave function renormalization of $M$ due to the superpotential interaction
$M q \overline q$. The sign of this correction is such that $M$ is stabilized at the origin.
As discussed in \cite{ISS} the theory also has supersymmetric vacua with large
expectation values for $M$. Thus the SUSY breaking vacuum is only metastable
but its lifetime is exponentially long in the parameter $m_A/m_Q$.

A nice feature of the ISS SUSY breaking model is that it
is robust against perturbations to the UV physics. We explained that
no new superpotential is generated when integrating out the field $A$. 
But a simple symmetry argument in the effective theory below the
mass of $A$ shows that even if additional terms were generated by
non-perturbative physics, they would not destabilize SUSY breaking.
The argument relies on the  $SU(F)\times SU(F)$ flavor symmetry
of the effective theory being only softly broken. 
The dynamically generated superpotential
must be a function of the $SU(F)\times SU(F)$
invariants $m^2 M$, $q \overline{q} M$, and det(M) or det$(q \overline{q})$.
Here $m^2$ is the properly normalized spurion of the IR which transforms
like $m_Q$ of the UV. Furthermore, the superpotential must be regular in the
IR fields $M$, $q$, $\overline{q}$, and therefore the most relevant terms
which are not already included in \refeq{dualsuperpot} are
$(m^2 M)^2/m_A$, $(M q \overline{q})^2/m_A^3$ or det$(M)/m_A^{F-3}$.
They are too small to destabilize the vacuum for $F > 3$. 
This feature makes the ISS model very attractive for our purposes. The
only significant constraint on UV modifications of the model is
that the $SU(F)\times SU(F)$ flavor symmetry is not strongly broken in the
conformal regime, as would happen if we were to
introduce a $\bar{Q} A Q$ superpotential.

In summary, we find that between $\Lambda_{\kappa}$ and $m_A$ 
the theory is approximately
conformal. Below $m_A$ the adjoint $A$ decouples and we are left with a 
strongly coupled non-conformal theory. In terms of the weakly coupled
dual variables, the IR dynamics is easy to understand;
SUSY is broken by the O'Raifeartaigh mechanism and a non-vanishing F-term for
some diagonal components of $M$ is generated.

We end this section by briefly outlining sequestering in our model, 
accounting for the fate of all operators of the form of
 Eq.~(\ref{thebadguy}), with fuller 
derivation of sequestering given in the following sections.  The only
$SU(N) \times SU(F)$ gauge-invariant operators of the form of
 Eq.~(\ref{thebadguy}) are  those corresponding to hidden bilinears
 $X^{\dagger} X = 
(\bar{Q}^{\dagger} \bar{Q} \pm Q^{\dagger} Q), A^{\dagger} A$.
Their dominant contribution to visible scalar masses can be determined by 
replacing the visible superfields $\Phi$ by their 
scalar components $\phi$ and treating these fields 
as spacetime constants, so that 
 Eq.~(\ref{thebadguy}) appears as a small $\phi$-dependent coupling 
constant multiplying $\int d^4 \theta X^{\dagger} X$. 
For the case $X^{\dagger} X = 
\bar{Q}^{\dagger} \bar{Q} - Q^{\dagger} Q$ this coupling can be 
field redefined away and therefore has no physical effect. 
For $X^{\dagger} X = 
(\bar{Q}^{\dagger} \bar{Q} + Q^{\dagger} Q), A^{\dagger} A$ these 
$\phi$-dependent couplings  cannot be redefined away completely,
because the redefinitions induce small $\phi$-dependent shifts of the $SU(N)$
gauge coupling (due to the Konishi anomaly \cite{Konishi}) and the $A^3$ 
Yukawa coupling away from their strong fixed-point values. 
The fact that these shifts are (technically) 
irrelevant physical couplings at the IR-attractive 
fixed point implies that the equivalent current operators, 
$\int d^4 \theta (\bar{Q}^{\dagger} \bar{Q} + Q^{\dagger} Q), \int d^4 \theta 
A^{\dagger} A$, are irrelevant too. Therefore $\phi$-dependence is
strongly suppressed in the IR.  This is conformal sequestering.

\section{Sequestering in detail}

In the introduction we 
recounted a general plan for conformal sequestering, 
based on the hidden sector dynamics spending a large hierarchy of energies 
in the vicinity of a strongly IR-attractive fixed point prior to 
SUSY breaking. However, there are three  subtleties one encounters 
in putting this plan to work: 

(i) The generic existence of relevant 
{\it superpotential} couplings in what are otherwise IR-attractive fixed points.

(ii) The generic existence of marginal operators associated to global 
symmetries at fixed points. 

(iii) We have thus-far neglected to  properly integrate out (quadratic) 
fluctuations in the 
visible $F_{\Phi}$-terms, rather than simply setting $F_{\Phi}$ to its 
vanishing VEV.

In this Section, we will first discuss issues (i) and (ii),  while continuing 
to make the ``mistake'' pointed out in (iii). 
 We will finally 
return to a proper treatment of issue (iii). It is important to work through 
all these issues in order to properly understand and check how the 
 model of Section 3
accomplishes the tasks of conformal sequestering followed by SUSY breaking. 
The model also illustrates all the 
general issues in a relatively simple setting.

\subsection{Relevant hidden superpotential couplings}

The first subtlety is the fact that typical ``IR-attractive''
 fixed points do indeed possess relevant perturbations, such as 
mass terms for some of the matter fields. A fixed point is usually deemed 
IR-attractive if such repulsive terms can forbidden by symmetries, as is the 
case for any relevant superpotential couplings, including supersymmetric 
masses. However, in the present context we cannot simply 
impose such symmetries, because we in fact need  
small mass terms  to ultimately drive the dynamics away from the fixed point 
towards SUSY breaking dynamics. According to our introduction we should then 
worry about terms in the Planck scale Lagrangian of the form, 
\begin{equation}
\Delta {\cal L}_{mixed} \sim \phi^{\dagger} \phi \int d^2 \theta X^n + 
{\rm h.c.},
\end{equation}
where $X^n$ represents some relevant hidden superpotential coupling. 
Fortunately though, such a term cannot arise from a full 
superspace invariant Lagrangian in terms of $\Phi$ and $X$  (where 
we only keep the lowest component $\phi$ of the visible superfield $\Phi$).

However, one can write unsequestered terms involving hidden 
superpotential couplings starting from mixed superpotentials,   
\begin{eqnarray}
\Delta {\cal L}_{mixed}(M) &\sim& 
\int d^2 \theta \frac{\Phi^k X^n}{M_{Pl}^{n + k -3}} + {\rm h.c.} \nonumber 
\\
&=& \frac{\phi^k}{M_{Pl}^{n + k -3}} 
\int d^2 \theta X^n + {\rm h.c.} + ...~ .  
\end{eqnarray}
In the IR the strong hidden scaling results in 
\begin{eqnarray}
\Delta {\cal L}_{mixed}(\mu) 
&\sim& (\frac{\mu}{M_{Pl}})^{\gamma} \frac{\phi^k}{M_{Pl}^{n + k -3}} 
\int d^2 \theta X^n + {\rm h.c.} + ...~   
\end{eqnarray}
where $\gamma$ is the anomalous dimension of $X^n$.
Upon SUSY breaking this results in visible sector A-terms. 
This issue can be avoided by simply making the standard plausible assumption
 that the superpotential at the Planck scale is 
sequestered without a protective symmetry, 
\begin{equation}
W(\Phi, X) = W_{vis}(\Phi) + W_{hid}(X),
\end{equation}
but protected by the non-renormalization theorem. 

Let us also consider the worst-case scenario in which 
 the mixed superpotentials are indeed present 
and estimate 
their visible effects.
Neglecting the 
thorny issue of the $\mu$ and $B \mu$ terms of the MSSM, let us  
focus on cubic visible gauge invariants, $k = 3$, and 
the resulting A-terms, 
\begin{equation}
A_{vis} \sim \Lambda_{int} (\frac{\Lambda_{int}}{M_{Pl}})^{n + \gamma}, 
\end{equation}
corresponding to a suppression of 
$(\frac{\Lambda_{int}}{M_{Pl}})^{n + \gamma - 1}$ over directly-mediated
 visible soft terms. This estimate is obtained by assuming that 
the fixed point behavior operates until the theory is not far above 
 $\Lambda_{int}$, and that this scale then sets 
all hidden sector VEVs.  
Note that $n + \gamma$ is the scaling dimension of $X^n$ at the fixed point, 
which is bounded by unitarity to be $\geq 1$, corresponding to suppression 
over direct-mediation as long as the composite, $X^n$, is interacting.

The R-symmetry of the superconformal algebra at the fixed point determines
the $\gamma$ of chiral operators. In our model,
there is a unique non-anomalous R-symmetry \cite{K} which shows that 
the most relevant chiral operator is $\bar{Q} Q$, with $\gamma = 1 - 2N/F$, 
corresponding to suppression of $A_{vis}$ relative to direct mediation of 
$(\Lambda_{int}/M_{Pl})^{2 - 2N/F}$. The danger is that this 
suppression disappears at the edge of the desired range, for $F \approx N$ 
where
$M=Q\overline Q$ becomes a free field.
Suppression of $(\frac{\Lambda_{int}}{M_{Pl}})^{1/2}$ is enough for A terms
and therefore requires $F \gsim 4N/3$. Of course, the real lesson here 
is that we should stay away from parameter choices where parts of the CFT 
are nearly free fields, such as $\bar{Q} Q$ for 
$F \approx N$, since strong coupling is the key to sequestering.

To conclude this subsection, the relevance of possible hidden 
superpotential couplings does not destroy the plot of 
conformal sequestering,  either because the Planck scale theory gives 
us a sequestered superpotential, or by ensuring that the entire 
hidden dynamics
is fully and strongly interacting at the fixed point.

\subsection{Marginal flavor-symmetry ``currents''}

The second subtlety is that at typical fixed points there are a set of 
global symmetries. These symmetries play an important role in the 
discovery of known strongly-coupled fixed points. The associated 
conserved Noether 
currents, $J_{\mu}^a$, 
(except for R-symmetries) are contained in 
supermultiplets of the form $X^{\dagger} T^a X$, where $T^a$ are
matrices representing  the symmetry generators
 acting on the $X$. Consequently, the standard
 vanishing of anomalous dimension 
for symmetry currents  translates by SUSY to the 
vanishing of anomalous dimension for the entire supermultiplets 
$X^{\dagger} T^a X$. That is, there is a bilinear $X^{\dagger} T^a X$ (with hidden gauge multiplets implicit as needed for gauge invariance) 
of dimension exactly 2 for every global symmetry at a fixed point. 
Therefore, terms in the Planck-scale Lagrangian of the form 
\begin{eqnarray}
\label{mixed}
\Delta {\cal L}_{mixed} &=& \frac{1}{M_{Pl}^2} 
\int d^4 \theta \Phi^{\dagger} \Phi 
X^{\dagger} T^a X \nonumber \\
&\rightarrow&  \frac{\phi^{\dagger} \phi }{M_{Pl}^2} 
\int d^4 \theta X^{\dagger} T^a X + ...~ , 
\end{eqnarray}
will not be conformally sequestered by the strong hidden fixed-point 
dynamics, indeed there is no running from this source at all!

We will use our hidden sector model to explain this loop-hole in 
our introductory argument guaranteeing sequestering, and to illustrate 
that while it poses the central danger in conformal sequestering, 
it is not fatal.
To identify the global 
symmetries of the strong conformal dynamics let us formally shut off all 
perturbations in the hidden sector, namely the weak $SU(F)$ gauge theory 
and the mass terms, $\alpha_{SU(F)}, m_A, m_Q \rightarrow 0$. The exact 
(non-R) symmetries are then those familiar from SQCD, namely 
$SU(F)_Q \times SU(F)_{\bar{Q}} \times U(1)_{baryon}$, 
with associated current multiplets $Q^{\dagger} T^a Q, ~
\bar{Q}^{\dagger} T^a \bar{Q}, 
Q^{\dagger} Q -  \bar{Q}^{\dagger} \bar{Q}$, 
where the $T^a$ span all traceless hermitian $F \times F$ matrices. These 
are dangerous because they have scaling (and canonical) dimension 2, 
corresponding to exactly marginal Kahler operators. The only other two $SU(N)$ 
gauge-invariant hidden-matter bilinears one can write are 
$Q^{\dagger} Q +  \bar{Q}^{\dagger} \bar{Q}$ and $A^{\dagger} A$. But these 
bilinears (or any linear combinations) do not correspond to symmetries 
because of the strong 
axial anomaly of the $SU(N)$ gauge theory and strong explicit 
breaking by the fixed point $A^3$ Yukawa coupling. Since we 
are at a strong IR attractive fixed-point the corresponding Kahler operators 
$\int d^4\! \theta\, Q^{\dagger} Q +  \bar{Q}^{\dagger} \bar{Q}, 
\int d^4\! \theta\, A^{\dagger} A$ are ${\cal O}(1)$ irrelevant and pose no 
danger. 

To understand what makes the conserved current bilinears special let us 
perturb the fixed point Lagrangian infinitesimally by the 
associated  Kahler terms.
We can do this by starting in the asymptotic
 UV (for this theoretical exercise, we 
are shutting off gravity and the visible sector) with
\begin{eqnarray}
\label{epsilon}
{\cal L}_{UV} &=& \int\! d^4 \theta\, Q^{\dagger} Q +  \bar{Q}^{\dagger} \bar{Q} + 
 A^{\dagger} A  + \int\! d^2 \theta\, \tau {\cal W}^2 + \kappa A^3 + {\rm h.c.} 
\nonumber \\
 &+& \int\! d^4 \theta\, \epsilon_a Q^{\dagger} T^a Q + 
\bar{\epsilon}_a \bar{Q}^{\dagger} T^a \bar{Q} + \epsilon 
(Q^{\dagger} Q -  \bar{Q}^{\dagger} \bar{Q})\ , 
\end{eqnarray}
where the $\epsilon$'s are infinitesimal. By the non-renormalization
of the conserved currents this
 flows in the far IR to 
\begin{eqnarray}
{\cal L}_{IR} &=& {\cal L}_{fixed-point} 
 + \int d^4 \theta \epsilon_a Q^{\dagger} T^a Q + 
\bar{\epsilon}_a \bar{Q}^{\dagger} T^a \bar{Q} + \epsilon 
(Q^{\dagger} Q -  \bar{Q}^{\dagger} \bar{Q})\ , 
\end{eqnarray}
which is what we want to study. 
However, note that in the UV the $\epsilon$ terms are merely deviations 
from canonical normalization for the matter fields. We can
return to canonical normalization by the field redefinitions,
\begin{eqnarray}
\eqn{transform}
Q &\rightarrow& (I - \frac{\epsilon}{2}- 
\frac{\epsilon_a}{2} T^a) Q \nonumber \\
\bar{Q} &\rightarrow& (I + \frac{\epsilon}{2} - 
\frac{\bar{\epsilon}_a}{2} T^a) \bar{Q}. 
\end{eqnarray}
These field transformations are related by SUSY to the infinitesimal 
symmetry transformations obtained by rotation of the $\epsilon$'s
in the complex plane, 
\begin{eqnarray}
Q &\rightarrow& (I - \frac{i \epsilon}{2}- 
\frac{i \epsilon_a}{2} T^a) Q \nonumber \\
\bar{Q} &\rightarrow& (I + \frac{ i \epsilon}{2} - 
\frac{i \bar{\epsilon}_a}{2} T^a) \bar{Q}, 
\end{eqnarray}
and as such, neither transformation is anomalous. It is straightforward to see 
that this transformation leaves the superpotential invariant.\footnote{In 
general the holomorphicity of superpotentials means that if they are invariant 
under symmetry group transformations with real 
parameters, $\epsilon$, they are 
automatically invariant under the corresponding complexified group 
transformations for complex $\epsilon$.} Note that one cannot similarly 
transform away infinitesimal  couplings of the form 
$Q^{\dagger} Q +  \bar{Q}^{\dagger} \bar{Q}$ and $A^{\dagger} A$, because 
the transformations are either anomalous or broken by the 
strong superpotential coupling $A^3$.

We thereby conclude that at the fixed point, the small change of wave-function 
normalization corresponding to the exact symmetry currents is physically 
irrelevant. That is, even though the local operator $\int\! d^4 \theta \,
X^{\dagger} T^a X(x)$ is a physical marginal operator at the fixed point, 
the associated Lagrangian term (zero momentum projection)
$\int d^4 x \int\! d^4 \theta\, X^{\dagger} T^a X(x)$ is not physical, and 
therefore does not represent a marginal coupling of the fixed point
(contradicting its being ``IR attractive'').


\subsection{Safe and unsafe currents}

Of course, we are really interested in mixed couplings like Eq.~(\ref{mixed}),
which are certainly physical since $\phi$ is in general a function of $x$. 
However, for the purpose of determining the visible masses alone we can 
treat 
$\phi^{\dagger} \phi$ as constant in spacetime. Thus, 
we can think of the various $\phi^{\dagger} \phi$ terms in  Eq.~(\ref{mixed}) as 
the $\epsilon$'s of Eq.~(\ref{epsilon}). If the hidden sector were only given by the 
fixed point dynamics, this would imply that the constant
$\phi^{\dagger} \phi$-dependence multiplying the symmetry currents can be 
completely transformed away as above, and no visible masses will result. 
However, the hidden sector also contains the perturbing couplings 
$\alpha_{SU(F)}, m_A, m_Q$.  We begin 
by turning back on the mass terms, $m_A, m_Q$, but leaving $SU(F)$ still 
ungauged, 
\begin{eqnarray}
{\cal L}_{UV} &=& \int\! d^4 \theta\, Q^{\dagger} Q +  \bar{Q}^{\dagger} \bar{Q} + 
 A^{\dagger} A  + \int\! d^2 \theta\, \tau {\cal W}^2 + \kappa A^3 + 
m_A A^2 + m_Q \bar{Q} Q + {\rm h.c.} 
\nonumber \\
 &+& \int\! d^4 \theta\, \epsilon_a Q^{\dagger} T^a Q + 
\bar{\epsilon}_a \bar{Q}^{\dagger} T^a \bar{Q} + \epsilon 
(Q^{\dagger} Q -  \bar{Q}^{\dagger} \bar{Q})\ , 
\end{eqnarray}
where the $\epsilon$'s represent $\phi^{\dagger} \phi$ constant 
visible bilinears. Now performing the transformation of Eq.~\refeq{transform} leads to 
\beq
\label{gone}
{\cal L}_{UV} &=& \int\! d^4 \theta\, Q^{\dagger} Q +  \bar{Q}^{\dagger} \bar{Q} + 
 A^{\dagger} A  \\ &+& \int\! d^2 \theta\, \tau {\cal W}^2 + \kappa A^3 + m_A A^2
+ m_Q \bar{Q} (I - \frac{\epsilon_a\!+\!\bar{\epsilon}_a}{2} T^a) Q + {\rm h.c.}
\nonumber 
\eeq
The visible terms are not transformed away, but rather multiply $m_Q$. 
Since the SUSY breaking vacuum energy of the isolated hidden sector is 
$V_{0} \propto |m_Q|^2$, in the presence of the
 visible sector perturbations, SUSY breaking leads to a potential
\begin{eqnarray} 
V_{eff} = V_0\ (1+ {\cal O}(\frac{\phi^{\dagger} \phi}{M_{Pl}^2}) ),
\end{eqnarray}
that is, unacceptable unsequestered visible scalar masses which dominate
over AMSB. We will deal with this in the next subsection. 

Note that the bilinear associated to $U(1)_{baryon}$ never had 
a chance to contribute to visible scalar masses because this symmetry 
is a symmetry of both the fixed point {\it and} the perturbations needed for 
SUSY breaking. As can be seen in Eq.~(\ref{gone}), the bilinear which couples
to the baryon number current is totally transformed away by Eq.~\refeq{transform}.

\subsection{Gauging flavor to suppress harmful non-abelian currents}

Let us turn back on our weak gauging of the vectorial $SU(F)$ 
symmetry. Since the current bilinears 
$Q^{\dagger} T^a Q,  ~  \bar{Q}^{\dagger} T^a \bar{Q}$ are adjoints of this 
weak gauge group, $SU(F)$ gauge-invariance forbids the mixed couplings of the form 
$\epsilon_a, \bar{\epsilon}_a$. Consequently, our model is indeed fully 
sequestered. 

\subsection{Strong superpotential coupling to suppress harmful abelian 
current}

Given our hidden field content, there is an alternative strong IR 
fixed point we might have thought to employ, namely the one arrived at 
by omitting the strong $A^3$ Yukawa coupling ($\kappa = 0$).
 However, this would have 
yielded one more (non-anomalous) 
$U(1)$ fixed point symmetry, corresponding to the 
bilinear $Q^{\dagger} Q +  \bar{Q}^{\dagger} \bar{Q} - \frac{F}{N} 
A^{\dagger} A$. It is straightforward to check that visible couplings
to this bilinear can be transformed away, but would result in 
 $\phi^{\dagger} \phi$ dependence multiplying both $m_A$ and $m_Q$, again 
resulting in a breakdown of sequestering. This $U(1)$ bilinear coupling is 
impossible to forbid by any weak gauging as in the previous subsection, 
because $U(1)$ currents are always 
singlets of any symmetry group, 
even the $U(1)$ itself. The strong $A^3$ Yukawa coupling 
was therefore crucial in having the resultant fixed point strongly break 
the $U(1)$ symmetry, turning the associated bilinear into an ${\cal O}(1)$ 
irrelevant coupling.

\subsection{Emergent symmetries?}

It should be stressed that the symmetries that pose a threat to conformal 
 sequestering are the global symmetries of the strong hidden dynamics 
at the fixed point. The strong fixed point is often arrived at theoretically 
by RG flowing from a weakly coupled theory in the UV. Necessarily, any 
global symmetry of the weakly coupled parent theory is also 
a symmetry of the strong IR fixed point. In our model, this symmetry is 
$SU(F)_Q \times SU(F)_{\bar{Q}} \times U(1)_{baryon}$, discussed above. 
However, in addition the strong 
IR fixed point may have ``emergent'' or accidental 
symmetries not present in the UV parent theory. 
While such  symmetries are equally dangerous to conformal sequestering,
their existence is clearly much more difficult to ascertain
because the weakly coupled UV parent cannot be consulted.
One approach is to plausibly conjecture that such emergent 
symmetries are simply absent at the IR fixed point. But when there are 
dual descriptions of such fixed points, these often offer a powerful 
 check of such a conjecture. This is the case in our model. 

The IR fixed point we employ has a dual description \cite{K,KS,KSS} in
terms of an $SU(2N\!-\! F)$ gauge theory with a superpotential
schematically of the form, 
\begin{equation}
W_{dual} \sim M \bar{q} a q +  N \bar{q} q + a^3,
\end{equation}
where $\bar{q}, q$ are F flavors of dual quarks, $a$ is a dual adjoint 
field and $M, N$ are gauge-singlet flavor-bifundamental meson fields. 
This dual parent theory has the same global symmetries as the 
original parent, namely $SU(F)_Q \times SU(F)_{\bar{Q}} \times U(1)_{baryon}$.
However, if $M \bar{q} a q$ was irrelevant in the IR, as suggested by 
canonical power-counting, then $M$ would decouple from the dynamics and
become a free field. In that case there would be an additional
$U(F^2)$ symmetry at the fixed point which transforms the decoupled fields $M$ 
freely among themselves. The true relevance of the 
$M \bar{q} a q$ superpotential coupling can be determined from
the superconformal R-charges and imposing the unitarity constraint that
any gauge invariant chiral primary field must have scaling dimensions greater
or equal to one. One finds that for $F \leq N$ this coupling must be
irrelevant because otherwise $M$'s dimensions would be less than one.
Thus the dual is allowing us to see the free field $M$ and the
associated emergent symmetries at the fixed point explicitly.
Fortunately however, the SUSY breaking mechanism of 
Ref.~\cite{ISS} operates for $F > N$, where one finds that the 
$M \bar{q} a q$ coupling is strong at the IR fixed point 
and as important as any of the other superpotential couplings. Thus, 
in our model, no new symmetries emerge in the dual description, and the 
conjecture that the only symmetries are 
$SU(F)_Q \times SU(F)_{\bar{Q}} \times U(1)_{baryon}$ is
strengthened.

An expert reader may have noticed that it was not really
necessary to consult the dual theory to determine if the gauge
invariant $M=Q\overline{Q}$ becomes a free field. Already in the
electric theory one can find the scaling dimension of $M$
from its superconformal R charge and see that its dimension
approaches 1 as $F$ approaches $N$ from above, suggesting
that $M$ is a free field for $F\leq N$.

However there is another worry regarding emergent symmetries
which only becomes apparent in the dual. The worry is that the $U(1)_A$
symmetry which we engineered to be broken explicitly by the $A^3$ term
might re-emerge at the fixed point. How might this happen?
Consider adding the $A^3$ term with small coefficient $\kappa$ to the
CFT with vanishing superpotential. At the $\kappa=0$ fixed point the dimension
of $A^3$ can be determined using R-charges and
a-maximization \cite{amax}. One finds that the operator $A^3$ is
relevant, driving the theory away from $\kappa=0$ towards 
large values of $\kappa$. This is good because it shows that our desired theory
(with $U(1)_A$  broken explicitly) is not unstable to flowing back
to the $\kappa=0$ fixed point, and we now have supporting evidence that the 
fixed point with $\kappa$ turned on strongly really does exist. What
about the dual? In the dual, the $U(1)_A$ symmetry is only broken
by the term $\tilde \kappa a^3$. Therefore, just as with $\kappa A^3$, 
we must worry that our desired fixed point with $U(1)_A$ broken
might not actually exist because it is unstable to flowing towards the
fixed point where $\tilde \kappa=0$. Happily, the superconformal
R-charges again determine that for small $\tilde \kappa$
the interaction $a^3$ is relevant, and therefore the theory also flows away
from this bad fixed point at which $U(1)_A$ is restored.
Altogether we have strong evidence that the desired fixed point with
no emergent symmetries exists and that it is stable.%
\ignore{

If this was the case,
then the hidden sector would have an ``emergent''
unsequestered dangerous symmetry. Looking in the dual, we see that the
$U(1)_A$ is now broken by the $a^3$ term.

 which is also a relevant 
operator. Thus $U(1)_A$ is also broken in the dual, giving
further evidence that $U(1)_A$ is not restored at the fixed point.
On the other hand, if we had found that $a^3$ in the dual is 
irrelevant, we would have had proof that $U(1)_A$ is a symmetry
of the fixed point.}%
\footnote{For an example where the dual provides evidence that
a symmetry similar to $U(1)_A$ emerges at a fixed point consider a slightly
different version of our model. Start with the ISS model and make it conformal
not by adding a massive adjoint $A$ and its superpotential $A^3$ but 
instead by adding $N/2$ extra massive flavors $P, \overline{P}$
with the superpotential $(P \overline{P})^2$. As in the case
of the adjoint, the superpotential was designed to break a
dangerous axial $U(1)_A$, and one can easily see from the R
symmetry that $(P \overline{P})^2$ is relevant (when added with small
coefficient) in the whole range of interest $N<F<3N/2$.
The R symmetry also shows that the meson $M=Q\overline Q$
becomes free for $F\leq 9N/8$. Therefore it seems that the model
should sequester as long as $9N/8<F<3N/2$.
However, in the Seiberg dual we discover a potential problem.
Like in the original variables $U(1)_A$ symmetry is broken by the superpotential
$\tilde \kappa (p \overline{p})^2$. However this time
we find that $ (p \overline{p})^2$ is actually irrelevant in the theory with $\tilde \kappa$
turned off for $F<1.24991 N$. In this regime of flavors 
the theory actually flows back towards $\tilde \kappa =0$ and $U(1)_A$ re-emerges.
This makes the desired fixed point at which $U(1)_A$ is strongly broken
highly suspicious because of the existence of an
alternative attractive fixed point nearby.
We conclude that this modified model appears to sequester safely
only for the relatively narrow range of flavors $1.24991 N < F < 1.5 N$.}

\subsection{Integrating out visible $F_{\Phi}$-terms}

Let us now take up point (iii) raised at the beginning of this Section, 
namely that until now we  have been ignoring  
visible auxiliary fields $F_{\Phi}$ within
mixed visible-hidden terms, 
\begin{eqnarray}
\label{Fvis}
\Delta {\cal L}_{mixed}(M) &\sim& \frac{1}{M_{Pl}^{n}} \int\! d^4 \theta \,
\Phi^{\dagger} \Phi\,  {\cal O}_{hid} \nonumber \\
\Delta {\cal L}_{mixed}(\mu) &\sim& 
\frac{\mu^{\gamma}}{M_{Pl}^{n + \gamma}} \int\! d^4 \theta \,
\Phi^{\dagger} \Phi\, {\cal O}_{hid},
\end{eqnarray}
where here ${\cal O}_{hid}$ denotes a (composite) hidden superfield with 
canonical dimension $n$ and fixed-point scaling dimension $n +  \gamma$. 
Assuming the fixed-point scaling holds down to energies modestly
larger than $\Lambda_{int}$, and
  integrating out the visible auxiliary field, results in visible scalar 
mass contributions, 
\begin{equation}
m^2_{\phi} \sim \frac{\Lambda_{int}^4}{M_{Pl}^2} 
(\frac{\Lambda_{int}}{M_{Pl}})^{2(n + \gamma -1)}.
\end{equation}
This is suppressed compared to direct mediation  contributions, by 
$(\Lambda_{int}/M_{Pl})^{2(n + \gamma -1)}$. By 
the unitarity constraints on conformal scaling dimensions, indeed 
$n + \gamma > 1$, as required for suppression. 

In general, we are unable 
to compute the strong anomalous dimensions $\gamma$, and  we rely on the 
fact that they are expected to be ${\cal O}(1)$ to provide 
sufficient suppression of this class of visible SUSY breaking 
in order for AMSB to dominate. There is however a special case when the 
hidden operator is chiral, 
${\cal O}_{hid} \sim X^n + $ h.c., for which the 
superconformal R-charge determines $\gamma$. Actually, 
 in this instance Eq.~(\ref{Fvis}) can be field redefined away via
\begin{equation}
\Phi \rightarrow \Phi(1 - \frac{X^n}{M_{Pl}^n}), 
\end{equation}
at the cost of leading to hidden contamination of the visible 
superpotential, 
\begin{equation}
W_{vis}(\Phi) \rightarrow W_{vis}(\Phi(1 - \frac{X^n}{M_{Pl}^n})).
\end{equation}
But the case of such
 visible-hidden mixed superpotentials has already been covered above.

\section{Discussion and Generalizations}

In this Section we collect and generalize our results and discuss
their robustness.
When analyzing candidates for models of hidden sectors with
dynamical SUSY breaking and conformal sequestering one must
perform a few consistency checks.

\bigskip

\noindent {\bf Consistency checks for dynamical SUSY breaking.}

\noindent

When integrating out heavy states at the scale of conformal
symmetry breaking one must check that dynamical superpotentials are not
generated which would otherwise ruin the IR SUSY breaking model.
We distinguish two cases, depending on whether conformal
symmetry breaking is explicit or spontaneous.

In the case of spontaneous breaking of conformal symmetry, there is a
Nambu-Goldstone boson, the ``dilaton''. The flatness of its potential
implies that no dynamical superpotential can be
generated. Instead explicit breaking by irrelevant operators must be 
used in order to generate a weak stabilizing dilaton potential \cite{LS1,LS2}.

The other case, explicit conformal symmetry breaking by a relevant operator,
as in this paper, is conceptually simpler. However, the relevant operator
necessarily strongly breaks the superconformal R-symmetry,
which is dangerous because R-symmetries are usually what prevent a
dynamical superpotential from being generated.
But in our model, the softly broken
$SU(F) \times SU(F)$ flavor symmetries of the
low energy effective theory (the ISS model) together with holomorphy
forbid any new superpotential couplings which
are relevant between the scales of conformal symmetry breaking and
SUSY breaking. Dynamically generated superpotential couplings which are
irrelevant between these scales do not destabilize SUSY breaking
because in our model their
effect on SUSY breaking can be made arbitrarily small, since the
hierarchy between conformal breaking and SUSY breaking is a free
parameter of the model.

Thus our SUSY breaking model is robust. In fact, it is easy to construct
other UV extensions of the ISS model which also exhibit conformal
sequestering. The only constraint from SUSY breaking is that the
$SU(F)\times SU(F)$ flavor symmetry be preserved in the conformal regime.

\bigskip

\noindent {\bf Consistency checks for conformal sequestering}

\noindent Here one needs to make sure that all possible couplings
between the hidden and visible sectors are sequestered.
As explained in this paper, the most dangerous couplings are relevant and
marginal operators of the CFT coupled to MSSM bilinears. The couplings to relevant
operators (which are gauge and superpotential couplings and therefore chiral)
can be forbidden by symmetries and are not a problem. More care is required in
studying possible marginal operators. In particular, for every global
symmetry of a CFT there exists a supermultiplet $X^\dagger T^a X$ which contains
the symmetry current and which is exactly marginal. Therefore operators of the form 
\beq
\frac{\Phi^\dagger \Phi}{M_{Pl}^2}\,  X^\dagger T^a X
\eqn{badoperator}
\eeq
are not renormalized by the conformal dynamics and do not sequester.
Sometimes these dangerous symmetries at a conformal fixed point are
difficult to spot because they are ``emergent'' or accidental as
discussed in subsection 4.6. Yet often they become apparent in dual
descriptions of the CFT.

To understand whether an operator like Eq.~\refeq{badoperator}
is problematic we distinguish four cases:

{\it i. Exact symmetries of the entire hidden sector.} In this case, 
 field transformations can be used to remove the 
couplings of the hidden currents to
 the MSSM bilinear, as discussed in Section 4.2 and illustrated by 
\refeq{transform}. From this
we conclude that even though the operator \refeq{badoperator} does not scale to
zero it does not give rise to scalar masses. This implies the following:

{\it If the action of a model of spontaneous SUSY breaking
has an exactly preserved global symmetry then the D component of the corresponding
supercurrent has a vanishing expectation value. i.e.
\beq
<\left. X^\dagger T^a e^V X\right|_D> = <F^\dagger T^a F> + <x^\dagger T^a D x> = 0
\eeq
where $x$ and $F$ are the scalar and F-components of $X$, and $D$ is a D-term
of hidden sector gauge fields. This is true even when the global symmetry is
spontaneously broken.}

{\it ii. Non-abelian symmetries.} The coupling of non-Abelian currents to
the visible sector can typically be forbidden by weakly gauging a subgroup
of the non-Abelian symmetry group.

{\it iii. Symmetries of the CFT dynamics which are broken by relevant operators
in the superpotential.} In this case the operator \refeq{badoperator} does not
sequester because while it contains a conserved current of the CFT 
 we cannot use the complexified symmetry  
transformations, such as Eq.~\refeq{transform}, to remove it because the
corresponding symmetry is broken. In this case sequestering fails and such
approximate symmetries must be avoided. In practice this means 
adding new interactions
to the CFT which cause it to flow to a new CFT where the symmetry is 
strongly broken. These
may be new strong gauge interactions or superpotential interactions
like the $\kappa A^3$ in our model. This leads to our next case below.

{\it iv. ``Symmetries'' which are strongly broken by the CFT dynamics.}
In this case the bilinear $X^\dagger T^a X$ does not correspond to a conserved
current because its associated symmetry is strongly broken by gauge
or superpotential couplings at the fixed point.
Since we are considering conformal fixed points which are IR attractive
the operator $X^\dagger T^a X$ must be irrelevant,
and the coupling \refeq{badoperator} scales to zero between the
Planck scale and the SUSY breaking scale. At low energies
it is suppressed by an additional
$(\Lambda_{int}/M_{Pl})^\gamma$ where the incalculable
anomalous dimension $\gamma$ is expected to be of order 1.

In conclusion, we have presented a class of hidden sector models
which exhibit conformal sequestering and break SUSY dynamically.
Sequestering is important as it is a necessary ingredient in models where SUSY 
breaking is mediated at a high scale (such as anomaly-, high scale gauge or gaugino-,
or graviton loop-mediation). We demonstrated that conformal sequestering
occurs in renormalizable four-dimensional models without relying on assumptions
about Planck scale physics or fortuitous discrete symmetries.
The simplicity and robustness of our models
leads us to believe that conformal sequestering is generic in the
``landscape'' of possible hidden sectors with dynamical SUSY breaking. 



\section{Acknowledgments}

We thank Andy Cohen, Paddy Fox, Ken Intriligator, Ami Katz, David Kaplan, Yuri Shirman,
Brian Wecht for useful discussions and the Aspen Center for Physics, the CERN
Theory Group, the Technion, and the Galileo Galilei Institute for their hospitality
while this research was nearing completion.
M.S. is supported by the DOE OJI grant DE-FG02-91ER40676 and an
Alfred P. Sloan Research Fellowship. R.S. is supported by NSF grant
P420-D36-2043-4350.

\appendix

\section{Appendix: SUSY QCD Sequestering}

The Lagrangian for supersymmetric QCD ($N$ colors, $F$ flavors) in the holomorphic basis is
\beq
\mathcal L_{hol} = Z \left. (Q_{hol}^\dagger Q_{hol}
+ \overline Q_{hol}^\dagger \overline Q_{hol})\!\right|_D
+ ( \tau \left. W_\alpha W^\alpha\right|_F + h.c. )
\eeq
where $\tau$ is the usual holomorphic coupling constant which runs at one
loop
\beq
\tau(\mu)=\tau(M)+\frac{b}{8\pi^2} \log(\frac{\mu}{M})\ , \quad\quad b=3N-F
\eeq
and the wave function $Z(\mu)$ gets contributions from all orders in perturbation
theory. In this basis, it is difficult to see how supersymmetric QCD could be 
conformal for any number of flavors except $F=3N$ because $\tau$ is manifestly
$\mu$ dependent. The resolution
of this puzzle lies in the running of the wave function of the matter fields:
Classically, the wave function factor $Z$ for the matter fields is unphysical, it can
be rescaled out of the Lagrangian by redefining fields $\hat Q = \sqrt{Z}\, Q_{hol}$,
$\hat {\overline Q} =  \sqrt{Z}\, \overline Q_{hol} $.
Quantum mechanically this transformation is anomalous \cite{Konishi}
and results in a shift of the coupling by $- \frac{F}{8 \pi^2} \log Z$.
Thus in the basis where the kinetic terms for the matter fields are canonical the 
running coupling is given by \cite{NSVZ,NSVZbynima}
\beq
\hat \tau(\mu)=\tau(M)+\frac{b}{8\pi^2} \log(\frac{\mu}{M}) - \frac{F}{8 \pi^2} \log Z(\mu)
\eeq
which is scale independent when the $\mu$ dependence of $Z$ is of the form
\beq
Z(\mu)=Z(M) \, \left(\frac{\mu}{M}\right)^{2\gamma_Q}\ , \quad\quad 2\gamma_Q=\frac{b}{F}=\frac{3N-F}{F}
\eeq

The basis in which both $Z$ and $\tau$ are running even though the
theory has only one physical coupling is confusing when
discussing conformal field theories. The two alternatives {\it i.} canonical
kinetic terms and all running in $\hat \tau$ and {\it ii.} fixing the gauge coupling by
moving the one-loop running of $\tau$ into the wave functions of the matter fields
are more convenient. In the following we give explicit formulae for the running 
couplings in all three bases. 

We start with the canonical basis in which the only running is in the gauge coupling.%
\footnote{Note that we use canonical kinetic terms for the matter fields but
not for the gauge fields.} 
The beta function expanded near the fixed point $\hat \tau = \tau_*$ is
\beq
\beta=0+{\beta'}_{\!*}\, (\hat\tau - \tau_*)
\eeq
with the solution
\beq
\hat \tau(\mu)-\tau_*=\left(\frac{\mu}{M}\right)^{{\beta'}_{\!*}} (\hat \tau(M)-\tau_*) \ .
\eeq
Thus the running Lagrangian in this basis is 
\beq
\hat{\mathcal{L}} = \left. (\hat Q^\dagger \hat Q + \hat{\bar Q}^\dagger \hat{\bar Q})\!\right|_D
+ ( \left[\tau_* + \left(\frac{\mu}{M}\right)^{{\beta'}_{\!*}} (\hat \tau(M)-\tau_*)\right] 
 \left. W_\alpha W^\alpha\right|_F  + h.c. )
\eeq
We now switch to the basis which we find most useful to discuss conformal sequestering,
the basis where all running takes place in the kinetic terms. The scaling of the kinetic
term is easily obtained by doing the general field redefinitions $\hat Q=\sqrt{R}\, Q$,
$\hat {\overline Q}=\sqrt{R}\, \overline Q$ under which
the gauge coupling shifts by $\frac{F}{8\pi^2} \log(R)$ and solving for $R$ such that the
new gauge coupling is fixed at $\tau_*$.
The result is
\beq
\mathcal L = \left[R(M)\right]^{(\frac{\mu}{M})^{{\beta'}_{\!*}}}
\left. (Q^\dagger Q + \bar Q^\dagger \bar Q)\!\right|_D
+ ( \tau_* \left. W_\alpha W^\alpha\right|_F  + h.c. )
\eqn{goodbasis}
\eeq
Note that the wave function factor rapidly approaches 1 as $\mu \rightarrow 0$.
For $R(M) \simeq 1$ it may be expanded to give
\beq
R(\mu)=1+\left(\frac{\mu}{M}\right)^{{\beta'}_{\!*}} (R(M)-1)
\eqn{QCDscaling}
\eeq
which is the formula we used earlier in \refeq{sqcdseq}.

For the purpose of conformal sequestering we are 
worried about operators of the form
\beq
c \frac{\Phi^\dagger \Phi}{M^2} \left. (Q^\dagger Q +\bar Q^\dagger \bar Q) \right|_D
\eeq
We may derive their scaling due to hidden sector interactions by
considering constant (scalar)
background values for the field $\Phi$ so that $c \frac{\Phi^\dagger \Phi}{M^2}$
becomes simply a number which contributes to the kinetic terms of $Q$ and $\bar Q$
i.e. it contributes a shift to $R(M)$.
But we know that all Lagrangian-dependence on $R(M)$ is suppressed by
a factor of $\left(\frac{\mu}{M}\right)^{{\beta'}_{\!*}}$,
therefore our operator must be sequestered
\beq
\mathcal O(\mu) = \left(\frac{\mu}{M}\right)^{{\beta'}_{\!*}}
c \frac{\Phi^\dagger \Phi}{M^2} \left. (Q^\dagger Q +\bar Q^\dagger \bar Q) \right|_D
\eeq

For completeness and to confuse you, we also give the running Lagrangian
in the holomorphic basis which is obtained from Eq.~\refeq{goodbasis} by moving the
one-loop running back into the gauge coupling
with the transformation $Q =  \left(\frac{\mu}{M} \right)^{\gamma_Q} Q_{hol}$.
The Lagrangian is then \cite{LS1}
\beq
\mathcal L_{ hol} = \left[R(M)\right]^{(\frac{\mu}{M})^{{\beta'}_{\!*}}}\!\!
&\!\!\left(\frac{\mu}{M}\right)^{2\gamma_Q}\!\! &\!\!
\left. (Q_{ hol}^\dagger Q_{ hol} + \bar Q_{hol}^\dagger \bar Q_{ hol})\!\right|_D\\
&+&\!\! ( \left[\tau_* + \frac{b}{8\pi^2} \log(\frac{\mu}{M}) \right]
\left. W_\alpha W^\alpha\right|_F + h.c. )
\eeq
Now it appears that there might be an extra suppression in conformal sequestering
due to the $\left(\frac{\mu}{M}\right)^{2\gamma_Q}$. But this factor
appears in front of the operator which couples visible and hidden sector as well as in
front of the kinetic terms for the $Q_{hol}$'s. Thus when we canonically
normalize hidden sector fields
this factor drops out again. This makes it clear that the anomalous scaling
dimensions at the fixed point $\gamma_Q$ do not contribute to sequestering.

\bibliography{seq}
\bibliographystyle{utcaps}

\end{document}